\documentclass[%
reprint,
prl,
 twocolumn,
]{revtex4-2}

\newcommand{\beqa}{\begin {eqnarray}}
\newcommand{\eeqa}{\end {eqnarray}}

\usepackage{natbib}
\usepackage{amsbsy}
\usepackage{amssymb}
\usepackage{amsmath}
\usepackage{graphicx}
\usepackage[justification=centering]{caption}
\usepackage{mathrsfs}
\usepackage[caption = false]{subfig}
\usepackage{array}
\usepackage{mathrsfs}
\usepackage{xfrac}
\usepackage{booktabs}
\usepackage{amsthm}
\usepackage{hyperref}
\hypersetup{colorlinks=true, citecolor=blue, urlcolor=blue, linkcolor=blue}

\usepackage{mathtools} 
\usepackage{graphicx}
\usepackage{dcolumn}
\usepackage{bm}
\usepackage{etoolbox}
\apptocmd{\sloppy}{\hbadness 10000\relax}{}{}

\usepackage{xcolor}
\usepackage{amsthm}
\usepackage{pifont}

\usepackage [english]{babel}
\usepackage [autostyle, english = american]{csquotes}
\usepackage{natbib}
\MakeOuterQuote{"}

\def\be{\begin{equation}}
\def\ee{\end{equation}}

\def\A{\mathcal {A}}

\def \nh{H~{\sc i~}}

\begin{document}
\title{Probing dark energy using anisotropies in the clustering of post-EoR \nh distribution}

\author{Chandrachud B.V. Dash}\email[E-mail: ]{cb.vaswar@gmail.com}

\author{Tapomoy Guha Sarkar}
\email[E-mail: ]{tapomoy1@gmail.com}

\affiliation{Department of Physics, Birla Institute of Technology and Science - Pilani, Rajasthan, India}

\begin{abstract}
We propose an anisotropy quantifier of the \nh 21-cm signal traditionally used to clock the astrophysics of the reionization era  as a post-reionization dark energy diagnostic. We find that the anisotropy probe can be measured at SNR $\sim 10$ in both auto-correlation and in cross-correlation with the Ly-$\alpha$ forest over a  wide $z$ and $k-$range. We propose to use the BAO signature on the anisotropy signal to measure $( H(z), D_A(z))$. Subsequently, we put constraints on a dark energy model involving a negative cosmological constant on top of a  quintessence scalar field and find that such a model is consistent with futuristic observations.

\end{abstract}

\maketitle

{\textit{Introduction:}}
An understanding of the nature of dark energy \cite{Padmanabhan_2003, RevModPhys.75.559} for explaining cosmic acceleration \cite{Perlmutter_1997, Spergel_2003, Hinshaw_2003, Riess_2016} remains elusive even today. The generally accepted $\Lambda$CDM paradigm also indicates tension with data. The measured value $H_0$ from high redshift CMBR observations  \cite{PhysRevD.104.083509, dutcher2021measurements,jain2003cross,huterer2002weak,amendola2008measuring,martinet2021probing} seems to consistently disagree with  low redshift estimates from  distance measurements using Supernova (SnIa) observations \cite{sandage2006hubble, riess2021cosmic, beaton2016carnegie, freedman2020calibration,blakeslee2021hubble} and from the imprint of baryon acoustic oscillation (BAO) signature in galaxy clustering \cite{seo2007improved}  or in the Ly-$\alpha$ forest \cite{Slosar_2009, baolyman2020}. Given the immense diversity of dark energy models spanning the theoretical landscape \cite{di2021realm}, and the critical observational crisis \cite{di2021snowmass2021,wong2020h0licow,abbott2022dark,verde2013planck,fields2011primordial},  it is imperative to measure the expansion history over a much wider redshift range and with greater level of precision.

Intensity mapping of the post-reionization HI distribution using the redshifted 21-cm signal holds the rich possibility of probing dark energy scenarios over a wide range of redshifts $0 \leq z \leq 6$ \cite{poreion0, poreion1, poreion2, poreion3, poreion4, poreion5, poreion6, poreion7, poreion8, poreion9, poreion10, poreion11, poreion12}. 
While it is theoretically promising, precise subtraction of large galactic and extra-galactic foregrounds \cite{Ghosh_2010, 2011MNRAS.418.2584G, liu2012well, liu2009improved,wang200621, di2002radio} 
remains the singularly crucial hindrance towards detecting the signal.
The cross-correlation of the  redshifted 21cm signal from the post-reionization epoch with other tracers from the same epoch has been extensively studied  as a probe of cosmological evolution  and structure formation \cite{Sarkar_2009, GSarkar_2010, Guha_Sarkar_2010, villaescusa2015cross,jelic2010cross, Sarkar_2015, Dash_2021,sarkar2016redshifted,Carucci_2017,dash2022probing}. 
The 21-cm foregrounds do not appear in the cross-correlation signal, and appears only as a noise. Therefore a statistically significant detection of the 21-cm signal can only be truly ascertained when detected in a cross-correlation \cite{chang2010hydrogen,amiri2023detection, amiri2024detection}. 

The redshifts probed by the 21-cm signal can also be probed by Ly-$\alpha$ forest, which is an established probe of the post-EoR epoch \cite{Croft_1999,Gnedin_2002, Viel_2002, Mandelbaum_2003, McDonald_2003,  coppolani2006, Gallerani_2006, McDonald_2007, D_Odorico_2006, Slosar_2009, McQuinn_2011,  Delubac_2015}. The Ly-$\alpha$ forest consists of distinct absorption features in the spectra of distant  QSOs and are tracers of the low column density HI density fluctuations along one dimensional sight lines. 
Observation of a large number of QSO spectra by the Baryon Oscillation Spectroscopic Survey (BOSS) \footnote{https://www.sdss3.org/surveys/boss.php}
with high signal to noise ratio (SNR) allows the possibility of a 
measurement of the $3D$ matter power spectrum \cite{McQuinn_2011} at redshift  $\sim 2.5$.
The cross-correlation of the Ly-$\alpha$ forest and the
redshifted 21cm signal as a powerful probe of the post-reionization epoch \cite{sarkar2013predictions, Sarkar_2019, Ashis-neutrino}  was proposed theoretically in linear theory \cite{Guha_Sarkar_2010, Sarkar_2015, sarkar2018predictions}, subsequently studied in numerical simulations \cite{Carucci_2017} and is recently detected in observations \cite{amiri2024detection}.

The 21-cm auto power spectrum and the 21-cm Ly-$\alpha$ cross-correlation power spectrum is anisotropic as it depends on $\mu = \cos \theta $ where, $\theta$ is the angle between the line of sight $\hat{n}$ and the Fourier mode vector $\mathbf{k}$.
This anisotropy arises from (i) the redshift space distortion (Kaiser effect \cite{kaiser1987clustering,hamilton1998linear}) arising due to the peculiar motion of the gas (ii) the nonlinear effect of peculiar velocity dispersion - {\it Finger of God effect} \cite{jackson1972critique} and (iii) the Alcock-Paczynski (AP) effect arising due to departures of distance scales from their values in a fiducial cosmology from their actual values along the line of sight and in the transverse direction \cite{AP1979,lopez2014alcock}.
In this paper we propose this anisotropy in the auto and cross-power spectrum to be a probe of dark energy.
As a suitable diagnostic for this anisotropy we use the quantity \cite{fialkov2015reconstructing}
\be  r(k, z) = \dfrac{\langle P( \mathbf{k}, z) \rangle_{|\mu| > 0.5} }{\langle  P( \mathbf{k}, z) \rangle_{|\mu| < 0.5} } - 1 , 
\label{eq:anisotropy}
\ee
where the average $\langle ~~ \rangle \equiv \int d \mu ~~ $ in the suitable $\mu$ range. 
This anisotropy ratio quantifies the $\mu$ dependence of the power spectrum and has values $r \sim 1$ for highly anisotropic power spectrum and is $r \sim 0$ when the $\mu$ dependence is negligible. This quantifier is used traditionally to "clock" the astrophysics of the re-ionization process \cite{fialkov2015reconstructing, ross2021redshift,sikder2024strong}. The anisotropy arises in the 21-cm signal from the cosmic-dawn  due to different powers of $\mu$  in the various terms of the  21-cm reionization power spectrum. 
In this paper we propose $r(k,z)$ as a dark energy diagnostic from the post-reionization epoch where the anisotropy is now rooted in the AP effect. The AP effect shall imprint the oscillatory BAO signature \cite{Hu-eisen} on $r(k,z)$ and can thus be used to constrain $D_A(z)$ and $H(z)$. The anisotropy ratio has the advantage that any improper modelling of the global signal including the neutral fraction does not affect $r(k,z)$. Further, the  differences in the anisotropy of the Ly-$\alpha$ forest signal from the 21-cm signal in the $(k_{\parallel}, k_{\perp})$ space  implies that the anisotropy ratio shall be higher for the cross-correlation. In this work we investigate $r(k,z)$ in the binned $(k,z)$ space as a means to optimally detect the BAO feature in both auto and cross signal and subsequently make error projections on various dark energy models. 

{\it Dark Energy Models: }
In this work we have considered the following dark energy scenarios governing  the low redshift cosmic evolution and structure formation.
Firstly we note that it is convenient to  use some parametrization of these models which mimic their general behaviour. The dynamic equation of state (EoS) $ p/\rho = w (z)$ is one such commonly used parametrization. The standard $\Lambda$CDM model is characterized by $w =-1$. In a two prameter model the EoS is given by $w(z) = w_0 + w_a  f(z)$.
The most popular and widely used parametrization  of this form is given by Chevallier-Linder-Polarski (CPL) \citep{CHEVALLIER_2001, PhysRevLett.90.091301}, where $f(z) = \dfrac{z}{1+z}$. This model gives a smooth variation of 
$w(z) = w_0 + w_a$ as $z\rightarrow \infty$ to $w(z) = w_0$ as $z \rightarrow 0$.
 A large  class of quintessence scalar field models can be mapped into the CPL parametrization \cite{scherrer2015mapping}.

Certain quintessence models with an  AdS vacuum   do not rule out the possibility of a negative cosmological constant $\Lambda$ \citep{dutta2020beyond, calderon2021negative, akarsu2020graduated, visinelli2019revisiting,ye2020hubble,yin2022small}. 
In this work we also consider such quintessence models  with a  non zero vacuum called CPL-$\Lambda$CDM models. 
These models mimic a rolling scalar field $\phi$ alongside a cosmological constant $\Lambda \neq 0$ \cite{sen2023cosmological, dash2024post}. 
In this work we assume that the Universe is governed by such a model with the best fit model parameters  ($\Omega_{m0} = 0.289 , \Omega_\Lambda = -0.781 ,  w_0 = -1.03, w_a = -0.1$) obtained from \cite{sen2023cosmological}.
We then use the BAO imprint on $r(k,z)$ to make error projections on $\Lambda$CDM, CPL, and CPL-$\Lambda$CDM  models.

{\bf \it The \nh 21-cm and Ly-$\alpha$ auto and cross-correlation: }
The astrophysical processes that characterized the EoR got over by $z \sim 6$ \cite{fan2002evolution} 
Most of \nh in the post-reionization era are housed in the Damped Ly-$\alpha$ (DLA) \cite{wolfe05} systems which source the HI 21-cm signal to be seen in emission against the background CMBR. A low resolution tomographic imaging of the diffuse HI 21-cm radiation background
without resolving  the individual DLAs is the aim of intensity mapping experiments. 
The diffuse  \nh in a predominantly ionized IGM in the same epoch also produces distinct absorption features,
in the spectra of background quasars. These Ly-$\alpha$ forest systems  traces \nh  density fluctuations along one dimensional quasar lines of sight. On large cosmological scales the Ly-$\alpha$
forest and the redshifted 21-cm signal are, both known to be biased tracers of
the underlying dark matter (DM) distribution \cite{McDonald_2003,Bagla_2010,Guha_Sarkar_2012,villaescusa2014modeling,Carucci_2017}.
The 3D cross-correlation power spectrum of  the smoothed fluctuations of the 21-cm brighness temperature $T$ and the  transmitted QSO flux $\cal{F}$ through the Ly-$\alpha$ forest as a function of $(z, {\bf k})$ is given by 
\begin{widetext}
\begin{eqnarray}
\label{eq:21cmps}
P_{ij}(k, z, \mu) &=& \frac{1}{\alpha_\parallel \alpha^2_\perp} \A_i \A_j  \left[ 1 + \beta_i \widetilde{\mu}^2  \right]  \left [ 1 + \beta_j \widetilde{\mu}^2 \right ]  P_m \left( \widetilde{k}, z \right)  \nonumber \\
{\rm where} ~~\alpha_{\parallel} &=& \frac{H^f}{H^r}~~\alpha_{\perp} = \frac{D_A^r(z)}{D_A^f(z) } ~~\widetilde{\mu}^2 = \frac{\mu^2}{F^2 + \mu^2 (1-F^2)} ~~\mathrm{and}~~ \widetilde{k} = \frac{k}{\alpha_\perp} \sqrt{1+\mu^2 (F^{-2}-1)} ~~{\rm with} ~~ F = \dfrac{\alpha_{\parallel}}{\alpha_{\perp}}
\end{eqnarray}
\end{widetext}
where $i,j \in (T, \cal {F})$.  We note that when $i = j = T$ we have the 21-cm auto power spectrum. Here $P_m(k, z)$ is the dark matter power spectrum at redshift $z$.
The amplitude of the 21-cm signal is given by \cite{poreion1,poreion2,poreion3}
\be \A_{T} = 4.0 \, {\rm {mK}} \,
b_{T}\, {\bar{x}_{\rm HI}}(1 + z)^2\left ( \dfrac{\Omega_{b0}
  h^2}{0.02} \right ) \left ( \dfrac{0.7}{h} \right) 
\dfrac{H_0}{H(z)} \ee
The mean neutral fraction $\bar{x}_{\rm HI} = 2.45 \times 10^{-2}$ \cite{xhibar1, xhibar2} is assumed to be a constant in the post EoR epoch. The 21-cm bias has been studied in \cite{Bagla_2010, Guha_Sarkar_2012, Sarkar_2016}. We have adopted the $(z,k)$ dependent bias $b_T$ fit function from \cite{Sarkar_2016} in this work. The redsift space distortion parameter $\beta_T(\widetilde{k}, z)$ for the 21-cm signal is given by $\beta_T = f/b_T$
where $f$ is the logarithmic growth rate of matter fluctuations. 
The amplitude of the Ly-$\alpha$ forest  transmitted flux fluctuation is adopted from the fitted flux data using BOSS \cite{font2018estimate}
$ A_\mathcal{F} = b_0   \left [  \dfrac{(1 + z)}{(1 + z_0)} \right ]^{2.9} $
 with fiducial $z_0 = 2.25$ and $b_0 = -0.14$ \cite{font2018estimate}. The corresponding distortion factor for  Ly-$\alpha$ forest is assumed to be  $\beta_F = 1.4$. 
If the fiducial cosmological model is different from the real Universe,  additional anisotropy shall be  introduced through the AP effect\citep{simpson2010difficulties,samushia2012interpreting,montanari2012new}.Thus, if $(H^f, H^r) $
and $(D_A^f, D_A^r)$ are the Hubble rate and angular diameter distance in the fiducial and real cosmology, additional AP anisotropy appears as in Eq.\ref{eq:21cmps} through the quantity $F$. 
\begin{figure*}[htbp]
\begin{center}
\includegraphics[scale=0.26]{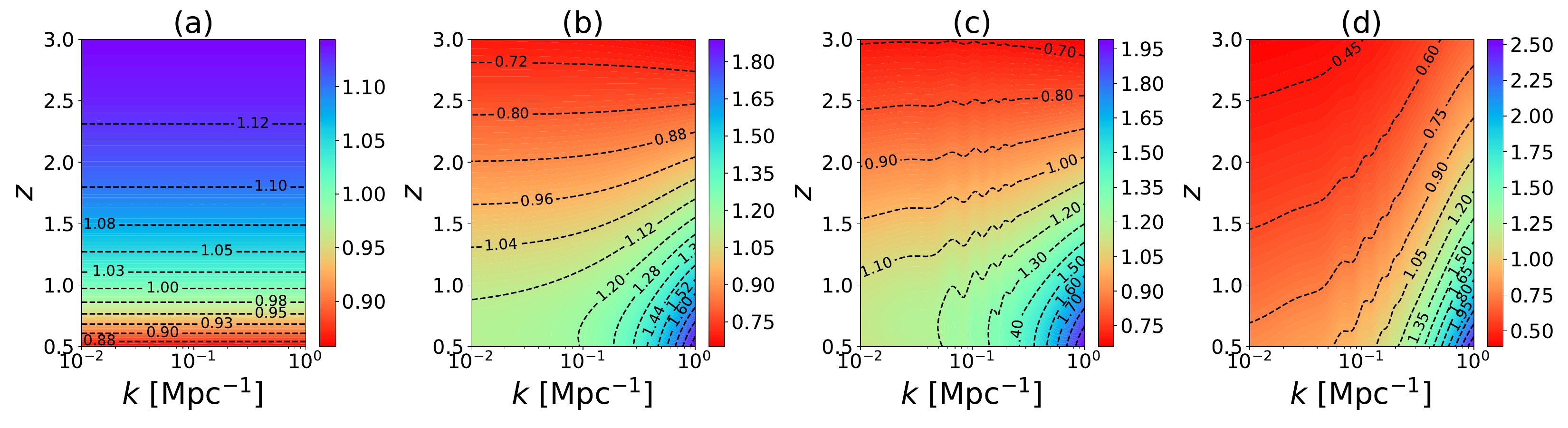}
\includegraphics[scale=0.26]{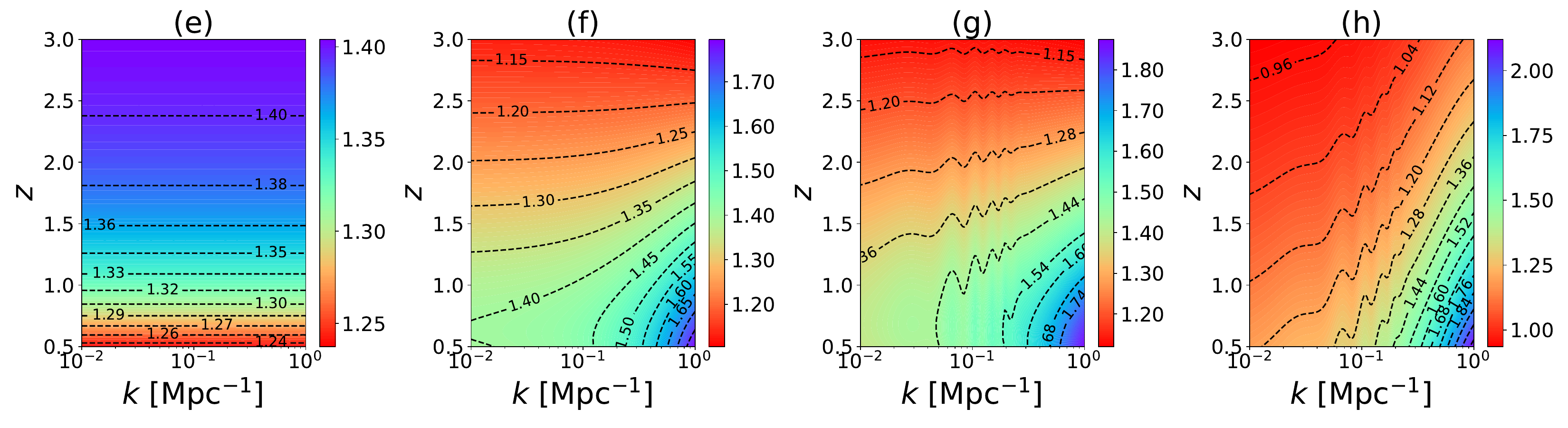}
\end{center}
\caption{$r(k,z)$ for auto-correlation (top) and cross-correlation with Lyman-$\alpha$ forest (bottom) in the $(k, z)$ space. Figure (a) is the case where $b_T ={\rm constant}$ (b) redshift and scale dependent $b_T(k,z)$ included (c) the AP effect included (d) AP effect along with a $k-$ dependent growth function. The lower panel (e)-(h) corresponding to the cross-correlation shares similar qualitative behaviour.}
\label{fig:anisotropy}
\end{figure*}

{\it Results and Discussion:}
This work focuses on the anisotropy diagnostic defined in Eq.(\ref{eq:anisotropy}) as a probe of cosmology in the post EoR epoch.
We first note  that in the absence of the AP effect, any overall function of $(k,z)$ shall not affect $r(k,z)$ which is determined by the relative coeffecients of the different powers of $\mu$. Thus, in this case the isotropic matter power spectrum $P_m(k,z)$ has no imprint on $r(k,z)$, the latter being only sensitive to the ratio  $\beta_T (k,z) = f/b_T$. In Fig.\ref{fig:anisotropy}(a) we consider a  constant linear bias $b_T={\rm constant}=1$. Here, since  $f(z)$ is $k-$independent, $r(k,z)$ is also scale independent and follows the $z$ dependence of $f(z)$. At large redshifts $f(z) \sim 1$ irrespective of dark energy models as perturbations grows as $\delta \propto a$ in the matter dominated epoch. Thus, in Fig.\ref{fig:anisotropy}(a) $r(z)$is seen to saturate at large redshifts. In this case $r(z)$ is thus a low redshift probe of $f(z)$. 

The assumption of constant bias $b_T$ is reasonable on large scales. However, in general $b_T(k,z)$ increases with $z$ monotonically and is $k-$dependent on small scales. We note that $r(k,z)$ is enhanced  when $\beta_T$ is large which corresponds to small $b_T$. This is reflected in Fig.\ref{fig:anisotropy}(b), where $r(k,z)$ is seen to decrease with $z$ since $b_T$ grows with $z$ faster than $f(z)$. As a function of $k$, the bias function for low redshifts is known to have a dip at around $k \sim 1 Mpc^{-1}$ \cite{Sarkar_2016}. The corresponding rise of $\beta_T$ with $k$ is  reflected in the peak in $r(k,z)$ at these scales for $z <1$. Thus, in models where $f$ is only a function of redshift, the scale dependence of $r$ thus gives a direct probe of the 21-cm bias without the issue of degeneracy with the overall amplitude of the power spectrum. 
The AP effect imprints $P_m(k,z)$ on $r(k,z)$ through the discrepancy in distance scales measured along the radial and transverse direction using the fiducial cosmology as compared to their real values. This introduces $\mu$ dependence in $P_m$ which manifests as the BAO oscillations in the $(k,z)$ plane as contours of $r(k,z)= {\rm constant}$ as is seen in Fig.\ref{fig:anisotropy}(c). In Fig.\ref{fig:anisotropy}(d), we consider a cosmological model with a Hu-Sawicki \cite{Hu_2007} $F(R)$ modification to gravity with $F_{,R0} = 10^{-5}$ \cite{Hu_2007, dash2020probing}. In such models, the growth rate is scale dependent and manifests in additional anisotropy on small scales at all redshifts. 
The lower panel in Fig.\ref{fig:anisotropy} shows $r(k,z)$ for the cross-correlation and demonstrates the same qualitative behaviour.
\begin{figure}[h]
\begin{center}
\includegraphics[scale=0.18]{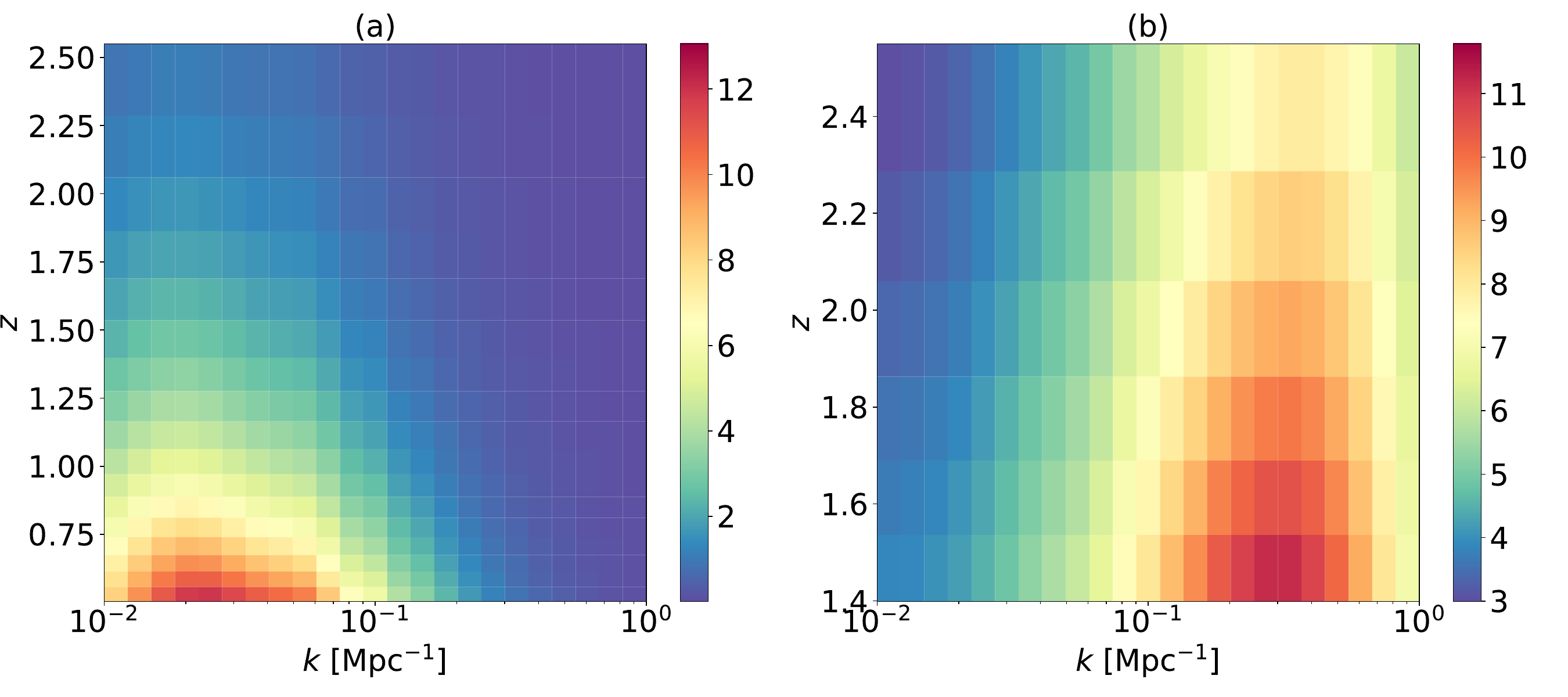}
\caption{SNR $ \left( \Delta r(k,z) /  r(k,z) \right)^{-1}$ for 21 cm auto-power spectrum (left) and for 21-cm  Ly-$\alpha$ cross power spectrum (right)}
\label{fig:SNR}
\end{center}
\end{figure}

We adopt a visibility-visibility correlation approach in a radio-interferometric observation  to make estimates for errors on the 21-cm auto power spectrum. We assume suitable averaging over visibility bins such that the covariance matrix is diagonal whereby \cite{mcquinn2006cosmological}
\begin{eqnarray}
\delta P_{TT}(k, \mu) = \frac{P_{TT} + N_T}{\sqrt{N_c}} 
~{\text {with}} ~ N_T = \frac{\lambda^2 T_{sys}^2 x^2 {dx}/{d\nu}  }{A_e t_{\bf k}} \nonumber
\end{eqnarray} 
Here for an observing redshift $z$, corresponding to a frequency $\nu$ we have $\lambda = 0.21(1+z)$m, $x$ is the comoving distance to the source, $A_e$ is the effective area of the antenna dish, $B$ is the bandwidth and 
\[ N_c = 2 \pi k^2 \Delta k \Delta \mu  x^2  ({dx}/{d\nu})B \lambda^2 / A_e (2 \pi)^3 .\]

The system temperature $T_{sys}$ is assumed to be dominated by sky temperature. For a total observation time $T_0$, we have $t_{\bf k} = T_o N_{ant} ( N_{ant} -1 ) A_e \rho / 2 \lambda^2$, where $N_{ant}$ is the number of antennae in the array and $\rho$ is the normalized baseline distribution function.

For the cross-correlation of 21-cm signal with the  Ly-$\alpha$ forest, the error is given by 
$$
\delta P_{T\cal F} ^2  = \frac{1}{2N_c} \left [ P_{T\cal F}^2 + \left ( P_{TT} + N_T \right ) \left ( P_{\cal F \cal F} +  N_{\cal F} \right ) \right ] $$
with, $N_{\cal F} = \frac{1}{\bar n} \int d{\bf k_{\perp}} P_{ \cal F \cal F}  + \frac{\sigma_F^2}{\bar{n}}$, where $\bar n$ is the quasar number density.

In order to make error projections on $r(k,z)$, we use the formalism in \cite{mcquinn2006cosmological}. Each term in $r(k,z)$  are integrals of $P(k, \mu,z)$ in some $\mu$ range. Thus $\sigma_r(k,z)^2/ r(k,z)^2$ is a sum in quadrature of the relative errors of each of those terms, which are of the form 
$\left (\int_{\mu -{\rm range} }  d\mu ~\delta P_{ij}(k, \mu)^{-2} \right )^{-1/2}\left (\int_{\mu- {\rm range} }  d\mu  P_{ij}(k, \mu)\right)^{-1}$.

For the radio-interferometric measurement of the 21-cm power spectrum we consider a futuristic
SKA1-Mid like experiment. We consider $250$ dish antennae each of diameter $15$m and efficiency $0.7$.  
The antennae are assumed to be distributed radially and falls off as $\sim 1/r^2$ which increases the noise on small scales due to poor visibility sampling. We consider about $17$ observations of bandwidth $32$ MHZ measurements at observing frequencies in the range $400$MHz to $950$MHz. We also consider $N_{\rm bin} = 24$ logarithmically binned $k$ values, in the range $0.005 {\rm Mpc}^{-1} \leq k \leq 1 {\rm Mpc}^{-1}$ with $\Delta k = \alpha k$ with $\alpha = 1/N_{\rm bin} \ln(k_{max}/ k_{min} )$. 
We consider a total observation time of $500$hrs per pointing and with multiple pointings for a full sky survey.
For cross correlation with the Ly-$\alpha$ forest we note that the quasar redshift distribution peaks in the range 
$2\leq z \leq 3$ \cite{Schneider_2005}. For quasars at redshift $\sim 3$, excluding some reshifts to avoid  quasar’s proximity
effect and contamination from Ly-$\beta$ forest or the intrinsic O-VI absorption, we focus on a redshift zone of $1.75 \leq z \leq 2.5$.
We also assume that all the Ly-$\alpha$ spectra are measured at a SNR of $6$ and consider a quasar number density of $\bar{n} = 32 \rm{deg}^{-2}$. The Ly-$\alpha$ and 21-cm data are also assumed to be smoothed to the same resolution before cross-correlating them.

Figure \ref{fig:SNR} shows $\Delta r (k,z) / r(k,z) $
in the $(k,z)$ plane. Fig.\ref{fig:SNR}(a) shows that using the 21-cm auto power spectrum 
$r(k,z)$ can be measured at $> 8 \sigma$ in the range $0.005 {\rm Mpc}^{-1} \leq k \leq 0.1  {\rm Mpc}^{-1} $ and $ 0.6 \leq z \leq 1.0$.  Fig.\ref{fig:SNR}(b) shows that $r(k,z)$ for the cross-correlation signal can be measured at $>8 \sigma$ sensitivity for $ {\rm Mpc}^{-1} 0.1 \leq k \leq 1 {\rm Mpc}^{-1} $.
\begin{figure}[h]
\centering
\begin{minipage}[b]{0.4\textwidth}
  \centering
  \includegraphics[scale=0.28]{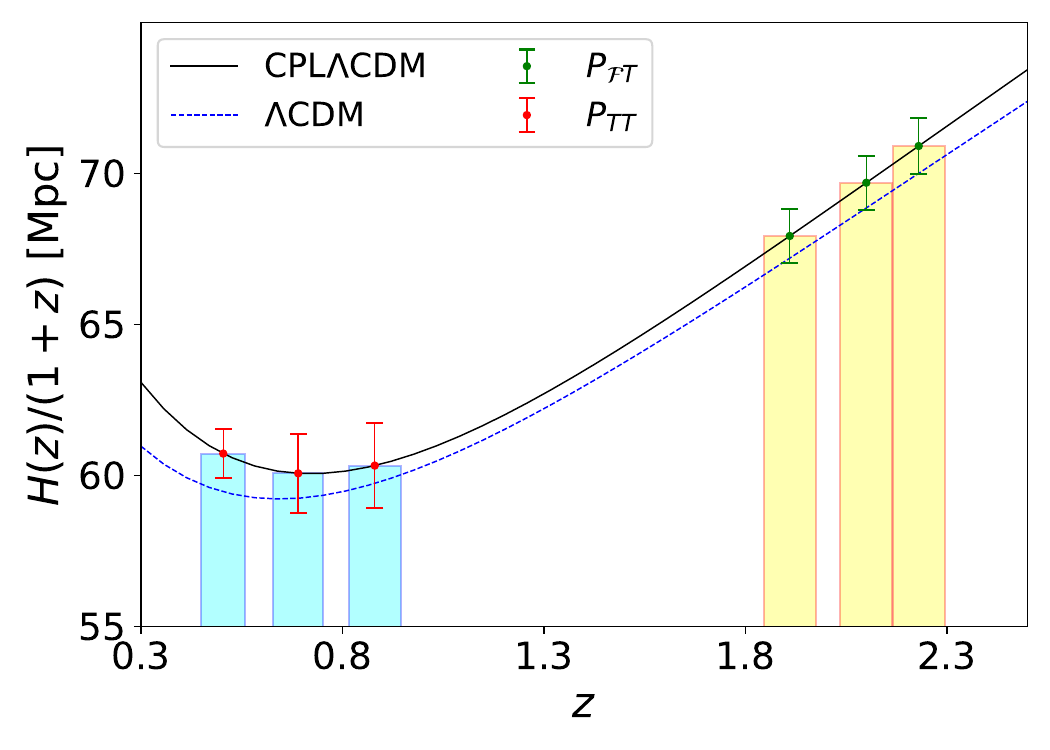}
\end{minipage}
\begin{minipage}[b]{0.4\textwidth}
  \centering
  \includegraphics[scale=0.28]{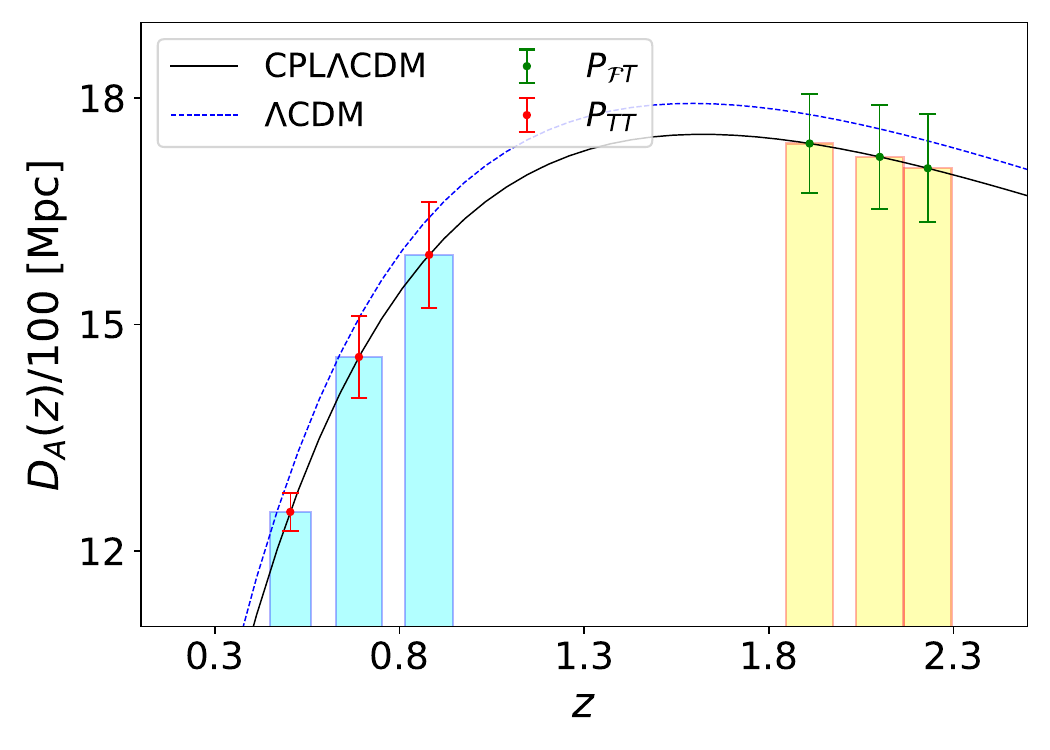}
\end{minipage}
\caption{shows the $H(z)$ and $D_A(z)$ error projections}
\label{fig:fisher-forecast}
\end{figure}

 The BAO scale $s(z_{d})$ is the sound horizon at the drag epoch $z_d$. We adopt the fiducial values  $z_{d} = 1060.01 \pm 0.29$ and $s (z_{d}) = 147.21 \pm 0.23 $Mpc from \citep{Planck2018} in our subsequent analysis. 
 The dependence of $r(k, z)$ on  BAO distance scale in the transverse and radial direction $(s_{\perp}, s_{\parallel})$ allows us to measure the angular diameter distance $D_A$ and Hubble rate $H(z)$.  
 The BAO feature in the matter power spectrum \cite{seo2007improved} is of the form $ \sim \dfrac {\sin \xi }{\xi} $, where 
$\xi = \sqrt {k_{\parallel}^2 s_{\parallel}^2  +  k_{\perp}^2 s_{\perp}^2  }$. We choose two parameters $q_1 = \ln s_{\perp}^{-1}$ and $q_2 = \ln s_{\parallel}$ so that $\delta q_1 = \delta D_A / D_A $ and $\delta q_2 = \delta H(z)/ H(z)$. The Fisher matrix  defined as 
\be F_{m n  } = \sum_k  \frac{1}{ \sigma_r(k,z)^2 }  \frac{\partial r(k,z) }{\partial q_{m}}     \frac{\partial r(k,z) }{\partial q_{n}}  
\ee
is used to obtain the $1-\sigma$ errors on the parameters $\delta q_m = \sqrt{ F^{-1}_{mm}}$. 
For this purpose we use the high SNR  regions identified  from Fig. \ref{fig:SNR}.  Figure \ref{fig:fisher-forecast} shows the 1-$\sigma$ errors on $D_A(z)$ and $H(z)$  for three z-bins at low redshifts obtained from the 21-cm auto correlation and three z-bins at high redshifts from the cross-correlation with Ly-$\alpha$.
\begin{figure}
\begin{center}
\includegraphics[scale=0.25]{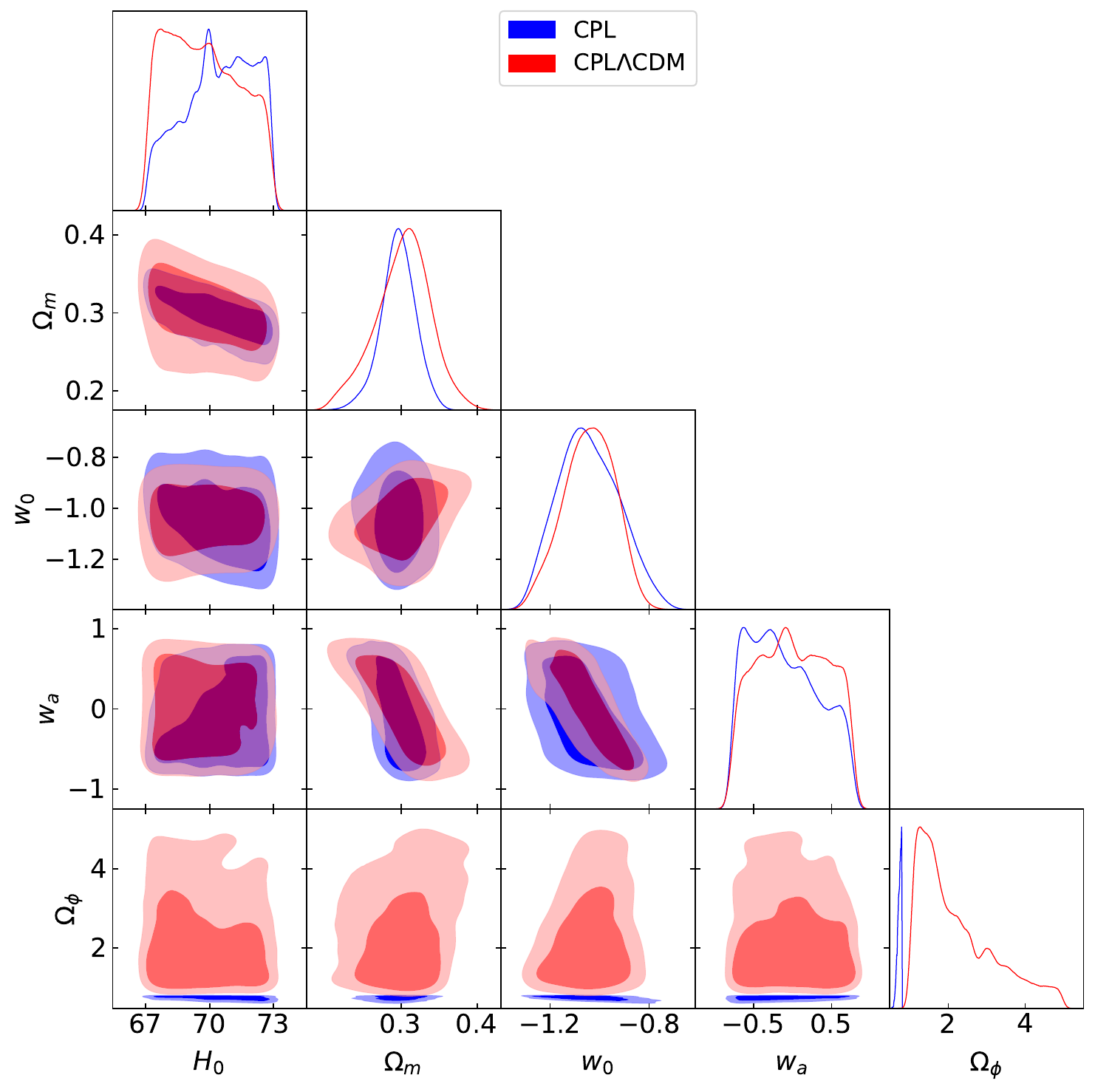}
\caption{Marginalized posterior distribution of the set of parameters ($H_0, \Omega_m,  w_0, w_a, \Omega_{\phi}$) and the corresponding 2D confidence contours obtained from the MCMC analysis. }
\label{fig:fisher-forecast-cpllcdm}
\end{center}
\end{figure}
\begin{table*}
\centering
\begin{tabular}{p{0.13\linewidth}p{0.12\linewidth}p{0.12\linewidth}p{0.12\linewidth}p{0.12\linewidth}p{0.12\linewidth}p{0.12\linewidth}}
\hline \hline
 Parameters &$H_0$ & $\Omega_{m}$  & $~~~~~~w_0$ & $~~~~w_a$ & ~~~~$\Omega_\phi$ \\ [0.5ex] 
 \hline\hline
Constraints \\ CPL-$\Lambda$CDM  & $69.624^{+2.070}_{-1.817}$ & $0.305^{+0.031}_{-0.040}$ & $-1.039^{+0.103}_{-0.109}$ & $-0.010^{+0.531}_{-0.488}$ & $1.914^{+1.402}_{-0.632}$ \\ 
 \hline
  ~~~~~~CPL  & $70.527^{+1.680}_{-1.944}$ & $0.296^{+0.022}_{-0.022}$  & $-1.052^{+0.141}_{-0.118}$ & $-0.148^{+0.574}_{-0.437}$ & $0.741^{+0.42}_{-0.059}$ \\
  \hline \hline
\end{tabular}
\caption{The parameter values obtained in the MCMC analysis are tabulated along with the $1-\sigma$ uncertainty.}
\label{tab:MCMC-constraints-cpllcdm}
\end{table*}
We find that a dark energy model with a scalar field along with  a negative cosmological constant and the $\Lambda$CDM can not be distinguished with these observations.

Taking flat priors for the CPL-$\Lambda$CDM model 
model parameters with ranges of $H_0 \in [67,73]$, $\Omega_m \in [0.2,0.6]$,  $\Omega_{\Lambda} \in [-7,2]$, $w_0 \in [-1.5, 1.5]$,   $w_a \in [-0.7, 0.7]$ we obtain the the marginalized posterior distribution of the set of parameters $(H_0, \Omega_m, \Omega_{\Lambda}, w_0, w_a)$  using a Monte Carlo Markov Chain (MCMC) analysis. Figure \ref{fig:fisher-forecast-cpllcdm} shows  the corresponding 2D confidence
contours.
The fiducial model parameters are taken from \cite{sen2023cosmological}. 
We also constrain the parameters for the CPL model with fiducial values adopted from \cite{hazra2015post}. The results are tabulated
in Table (\ref{tab:MCMC-constraints-cpllcdm}). 
We find  that even high redshift measurements of $H_0$ in $ 0.5 \leq z \leq 2.5$ using the proposed anisotropy diagnostic indicates that  dark energy with AdS vacua is consistent with cosmological
observation including the local R21 measurements of $H_0$ \cite{riess2022comprehensive}.

We conclude, by noting that $r(k,z)$ is a potentially powerful  probes of dark energy models at redshifts where galaxies and CMBR can not probe.
Further, though $r(k,z)$ has a similar appearance as the spherically averaged power spectrum, it has two distinct advantages. Firstly it is a ratio and is hence insensitive to the global signal and does not require a  precise measurement of $x_{HI}$. Secondly we note that foreground modeling and subtraction is key to detecting the 21-cm signal \cite{Ghosh_2010, 2011MNRAS.418.2584G, liu2012well, liu2009improved, wang200621, Liu-formalism1}. We know that for a radio field of view $\theta_o$  one has to avoid the foreground wedge \cite{Liu-formalism1} $|\mu| < \alpha / \sqrt{1 + \alpha^2}$ where $ \alpha = H(z) D_A(z) \theta_o/c$. 
$r(k,z)$ subtracts out the $\mu$ integral for small $\mu$s in the numerator and is thus is enhanced when the low $\mu$ modes are eliminated. A detailed treatment of the issues of foregrounds and other systematics would require 
a more comprehensive work.

\bibliographystyle{apsrev4-2}
\bibliography{references}
\end{document}